# Adaptive Frameless Rendering

Abhinav Dayal[1], Cliff Woolley[2], Benjamin Watson[1] and David Luebke[2]
[1]Northwestern University, [2]University of Virginia

**Abstract**
*We propose an adaptive form of frameless rendering with the potential to dramatically increase rendering speed over conventional interactive rendering approaches. Without the rigid sampling patterns of framed renderers, sampling and reconstruction can adapt with very fine granularity to spatio-temporal color change. A sampler uses closed-loop feedback to guide sampling toward edges or motion in the image. Temporally deep buffers store all the samples created over a short time interval for use in reconstruction and as sampler feedback. GPU-based reconstruction responds both to sampling density and space-time color gradients. Where the displayed scene is static, spatial color change dominates and older samples are given significant weight in reconstruction, resulting in sharper and eventually antialiased images. Where the scene is dynamic, more recent samples are emphasized, resulting in less sharp but more up-to-date images. We also use sample reprojection to improve reconstruction and guide sampling toward occlusion edges, undersampled regions, and specular highlights. In simulation our frameless renderer requires an order of magnitude fewer samples than traditional rendering of similar visual quality (as measured by RMS error), while introducing overhead amounting to 15% of computation time.*

Categories and Subject Descriptors: I.3.3 [Computer Graphics]: Picture-Image Generation—Display algorithms; I.3.7 [Computer Graphics]: Three-Dimensional Graphics And Realism—Raytracing; Virtual reality

## 1. Improving Interactive Rendering

In recent years a number of traditionally offline rendering algorithms have become interactive or nearly so. The introduction of programmable high-precision graphics processors (GPUs) has drastically expanded the range of algorithms that can be employed in real-time graphics; meanwhile, the steady progress of Moore's Law has made techniques such as ray tracing, long considered a slow algorithm suited only for offline realistic rendering, feasible in real-time rendering settings. These trends are related; indeed, some interactive global illumination research performs algorithms such as ray tracing and photon mapping directly on the GPU [PBMH02]. Future hardware should provide even better support for these algorithms, bringing us closer to the day when ray-based algorithms are an accepted and powerful component of every interactive rendering system.

What makes interactive ray tracing attractive? Researchers in the area have commented on ray tracing's ability to model physically accurate global illumination phenomena, its easy applicability to different shaders and primitives, and its output-sensitive running time, which is only weakly dependent on scene complexity [WPS*03].

We focus on another unique capability: selective sampling of the image plane. By design, depth-buffered rasterization must generate an entire image at a given time, but ray-tracing can focus rendering with very fine granularity. This ability enables a new approach to rendering that is both more interactive and more accurate.

The topic of sampling in ray tracing may seem nearly exhausted, but most previous work has focused on *spatial*

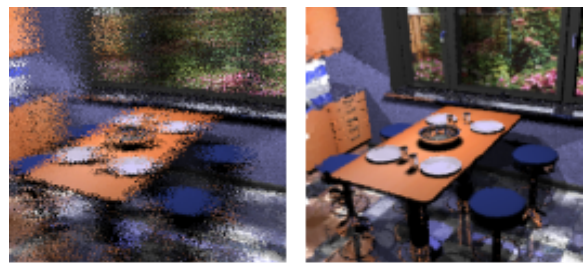

**Figure 1:** *Adaptive frameless rendering improves upon frameless rendering [BFMS94] (left) with adaptive sampling and reconstruction (right). Resulting imagery has similar visual quality to a framed renderer but is produced using an order of magnitude fewer samples per second.*

*sampling*, or where to sample in the image plane. In an interactive setting, the question of *temporal sampling*, or when to sample with respect to user input, becomes equally important. Temporal sampling in traditional graphics is bound to the frame: an image is begun in the back buffer incorporating the latest user input, but by the time the frame is swapped to the front buffer for display, the image reflects stale input. To mitigate this, interactive rendering systems typically increase the frame rate by reducing the complexity of the scene, trading off fidelity for performance. A few systems (e.g., [WDP99; SS00; TPWG02]) have explored an alternative: reuse samples over multiple frames, extending the useful life of prior rendering computations. Our own work falls into this second category.

Specifically, we investigate novel sampling schemes for managing the fidelity-performance tradeoff. Our approach has three important implications. First, we advocate *adaptive spatio-temporal sampling*, analogous to the adaptive spatial sampling long employed in progressive ray tracing [BFGS86; M87; PS89]. Spatially adaptive renderers have long focused rendering computation *where* it is most important; spatio-temporally adaptive sampling also focuses computation *when* it is most important. Second, we advocate *frameless rendering* [BFMS94], in which samples are located freely in space-time rather than placed at regular temporal intervals forming frames. Frameless rendering provides lower latency than framed rendering since every sample reflects the most recent user input. Third, we advocate *temporally adaptive reconstruction*, in which images are reconstructed from a sampled space-time volume, rather than a coherent temporal slice. We observe that the importance of old samples during this reconstruction varies according to the local temporal gradient, and adjust our reconstruction filter accordingly.

Our prototype adaptive frameless renderer consists of two subsystems. An *adaptive sampler* directs rendering to image regions undergoing significant change (across space and/or time). The sampler produces a stream of samples scattered across space-time; recent samples are collected and stored in two temporally deep buffers. One of these buffers provides feedback to the sampler, while the other serves as input to an *adaptive reconstructor,* which repeatedly reconstructs the samples in its deep buffer into an image for display, adapting the reconstruction filters to local sampling density and color gradients. Where the displayed scene is static, spatial color change dominates and older samples are given significant weight in reconstruction, resulting in sharper images. Where the scene is dynamic, only more recent samples are emphasized, resulting in a less sharp but correctly up-to-date image.

We describe an interactive system built on these principles, and show in simulation that this system achieves superior rendering accuracy and responsiveness. We compare our system's imagery to the imagery that would be displayed by a hypothetical zero-delay, antialiased renderer using RMS error. Our system outperforms not only frameless sampling (Figure 1), but also equals the performance of a framed renderer sampling 10 times more quickly.

## 1.1. Contributions

Our work introduces several new ideas to the existing body of research on sample reuse:

- An incrementally updated tiling that guides adaptive frameless sampling by spatio-temporal color variation.
- Use of methods from closed-loop control to improve adaptive sampling quality.
- Estimation of color gradients in time, using the "crosshair" mechanism described in Section 4.1.
- Use of filter kernels in image reconstruction that span not just space, but also time.
- Use of temporal gradients to shape these filter kernels, permitting low latency response or antialiasing, depending on scene content.
- Decoupling of the structures (the "deep buffers") used to guide adaptive sampling and image reconstruction/display.
- A GPU implementation of point-based reconstruction with support for arbitrarily sized space-time filters.

Ultimately, we argue that (with these improvements) the use of framelessly distributed samples can reduce latency and improve adaptivity in interactive systems, and has the potential to reduce synchronization requirements in parallel computer graphics.

## 2. Related work

Bishop et al.'s *frameless rendering* [BFMS94] replaces the coherent, simultaneous, double-buffered update of all pixels with samples distributed stochastically in space, each representing the most current input when the sample was taken. Pixels in a frameless image therefore represent many moments in time. Resulting images are more up-to-date than double-buffered frames, but temporal incoherence causes visual artifacts in dynamic scenes. A later technical report [S97] proposed (but did not implement) several extensions; adaptive frameless rendering incorporates some of these.

Other researchers have also examined loosening framed sampling constraints. Just-in-time pixels [OCMB95] takes a new temporal sample for each scanline. Interruptible rendering [WLWD03] uses a temporally adaptive framed sampling scheme that adaptively controls frame rate to minimize simultaneously the error created by reduced rendering fidelity (coarse imagery) and by reduced rendering update (late imagery). The address recalculation pipeline [RP94] sorts objects into several layered frame buffers refreshed at different rates. Talisman [TK96] renders portions of the 3D scene at different rates. Corrective texturing [SHSS00] uses rasterization hardware to generate images rapidly, and lazily fills in details from an ongoing global illumination simulation as progressively refined superimposed texture maps. Ward and Simmons [WS99] and Bala et al. [BDT99] store and reuse previously rendered rays. Havran et al. [HDM03] calculate the temporal interval over which a given sample will remain visible in an offline animation and reproject that sample during the interval, recalculating shading for all reprojected samples in every frame.

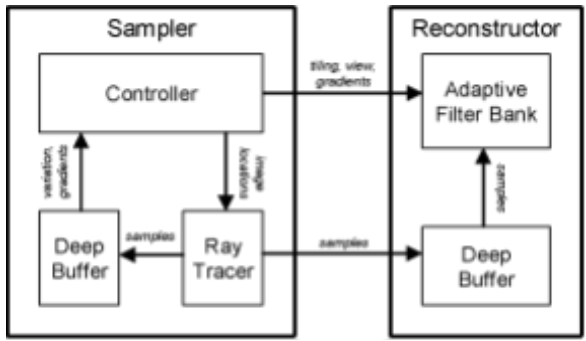

**Figure 2:** *Adaptive frameless rendering system components and data flow.*

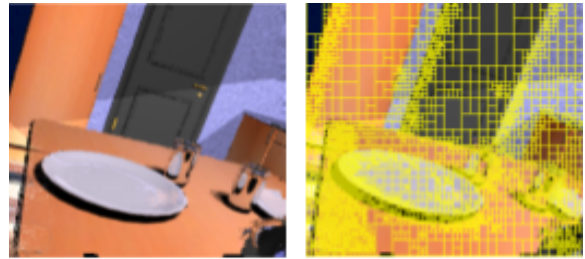

**Figure 3:** *A reconstructed image and an overlay showing the tiling used by the sampler at that moment in time. Note the finer tilings over object edges and occlusions.*

Several researchers have studied *sample reprojection*, which reuses samples from previous frames by repositioning them to reflect the current viewpoint. Some of this work is particularly relevant, and we make detailed comparisons to three important examples. The *Render Cache* by Walter et al. [WDP99; WDG02] is most closely related to our work; it reprojects samples each frame to account for camera motion and applies an image-space filter to reconstruct. The *Tapestry* system by Simmons and Séquin [SS00], and the *Shading Cache* by Tolé et al. [TPWG02] both integrate new samples into 3D meshes to use hardware-accelerated projection and Gouraud shading for image reconstruction. All three techniques perform prioritized sampling with similar goals to our own. We provide detailed comparisons after describing our own sampling and reconstruction techniques in Sections 4 and 5.

Many recent advances have made high-speed ray tracing a reality. These include clever optimizations and improved memory locality [PKGH97; TA98; WBWS01] as well as advances in the underlying hardware. Researchers have demonstrated interactive ray tracers on supercomputers [PMS*99], on PC clusters [WSB01; WBDS03], on the SIMD instruction sets of modern CPUs [WBWS01; RSH05], on graphics hardware [PBMH02; CHH02], and on custom hardware [SWWPS04]. These advances will soon make interactive ray-based rendering commonplace, enabling fine-grained selective sampling in real time.

### 3. Algorithm overview

Our system has two major components (Figure 2). The *sampler* consists of a controller, a ray tracer guided by the controller, and a temporally *deep buffer* that stores samples for feedback to the controller. For implementation efficiency, the *reconstructor* keeps a second deep buffer. The samples in this buffer are the input to the reconstructor's adaptive filter bank, which forms images according to local estimates of color gradients and sample density.

The sampler strives to increase sampling frequency in image regions where color variation across space and/or time is high. Because every new frameless sample is more current than the last, local sample density varies not only across space (for spatially adaptive response to edges), but also across time (for temporally adaptive response to motion). To track color variation, the controller uses an image-space *tiling* (Figure 3) of the deep buffer. In this frameless context, sampled content is constantly changing. The controller therefore continually adjusts the tiling using merges and splits, ensuring that tiles cover roughly equal amounts of color variation, with small tiles located over scene edges and motion. To decide where to sample next, the controller simply picks a tile at random. While adding new samples to a tile, the controller also reprojects several samples covered by the tile to new deep buffer locations that reflect the current view. This improves the feedback provided by the deep buffer to the controller and permits improved sampling response to occlusion.

The reconstructor strives to provide spatially detailed, antialiased imagery where the scene is static, and low-latency (if blurred) imagery where the scene is dynamic. It achieves this with locally adaptive space-time filtering (Figure 6). The sampler streams the same samples placed in its own deep buffer to the reconstructor's deep buffer. At each display refresh, the sampler also provides the reconstructor with local color gradient and sample density information. This information is then used to "shape" and "size" local space-time filters in the reconstructor's adaptive filter bank. Where scene content is static, temporal color gradients will be low, and temporal filter extents can incorporate older buffer samples for antialiasing. Where scene content is dynamic, temporal gradients will be high, and temporal filter extents can ignore older buffer samples for low-latency response. To improve resulting imagery, the reconstructor reprojects all of the samples in its buffer to reflect current view information. We have implemented the entire reconstructor on the GPU.

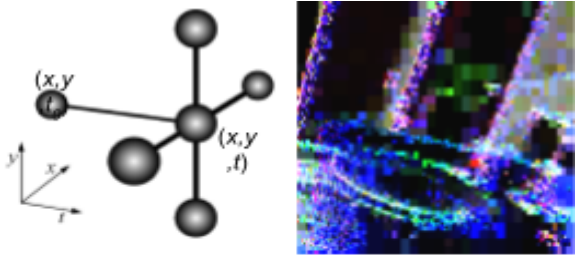

**Figure 4**: *We organize samples into "crosshairs" (left) to compute spatial and temporal tile gradients (right), shown here as red ($G_x$), green ($G_y$), and blue ($G_t$).*

## 4. Adaptive frameless sampling

Because sampling is frameless and distributed across time as well as space, samples are collected not into a frame but into a temporally deep buffer. This deep buffer is a 3D array sized to match the number of image pixels in two dimensions, and to accommodate a shallow buffer depth *b* in the temporal dimension (we use $b = 4$). Buffer entries at each pixel location form a queue, with new samples inserted into the front causing the removal of samples in the back if the queue is full. Each sample is described by its color, position in world space, age, and optionally a view-independent velocity vector.

Like previous importance sampling techniques [BFGS86; G95; M87; PS89], our controller uses an image-space tiling to guide adaptive sampling. However, while these previous techniques worked in a static framed context, our sampler operates in a dynamic, frameless context. The controller's tiles therefore partition not just image space, but also segment the deep buffer into space-time volumes called *blocks* using planes parallel to the temporal axis. To remain useful, this tiling must constantly change in response to user interaction and animation. We manage this change using a K-D tree, with the current tiling implemented as a cut across the tree. Given a target number of tiles, the tree is managed to ensure that the amount of color variation in each tile's block is roughly equal: the tile with the most color variation is split and the two tiles with the least summed variation are merged, until all tiles have roughly equal variation. As a result, small tiles emerge over image regions containing edges and motion (Figure 3).

We calculate spatio-temporal color variation within a block using the equation $v_{tile} = 1/n \, \Sigma_i \, (L_i - L_m)^2$, where $L_i$ is a sample's luminance and $L_m$ the mean luminance in the block. We ensure prompt response to changes in scene content by weighting samples in the variance calculation using a function that declines exponentially as sample age increases: $e^{-3.47a}$, where *a* is sample age.

Sampling is a biased probabilistic process that makes use of these tools for measuring color distribution. Since the current time is not fixed as it would be in a framed renderer, we cannot simply iteratively sample the tile with the most variation—in doing so, we would overlook newly emerging motion and detail. At the same time, we cannot leave rendering non-adaptive and unimproved. Our solution is to iteratively select the next tile to sample randomly using a uniform distribution, and to choose the sampled location within the selected tile similarly. Because tiles vary in size, sampling is adaptively biased towards those regions of the image which exhibit high spatial and/or temporal color variation. Because all tiles are randomly sampled, we remain sensitive to newly emerging motion and detail.

### 4.1. Gain control and gradient sampling

Our adaptive sampler's controller is less effective when the rendered scene is more dynamic, making the color distribution more difficult to track. To address this problem we apply a control engineering technique: adjusting *gain*. We implement this by adjusting the number of tiles onscreen, restricting the ability of the sampler to adapt to deep buffer content when the scene is dynamic, and increasing this ability when the scene is static. Specifically, we change the target number of image-space tiles so that color change over space and time are roughly equal in all blocks by ensuring that $dC/ds \, S = i \, dC/dt \, T$, where $dC/ds$ and $dC/dt$ are spatial and temporal color gradients averaged over the entire image (Figure 4), *S* is the average width of the tiles, *T* the average age of the samples in each tile, and *i* is a constant adjusting the relative importance of temporal and spatial color change in control. By solving for *S* we can derive the appropriate number of tiles.

To sample color gradients, we organize all samples into spatio-temporal *crosshairs* (Figure 4), each of which forms a single entry in the sampler's deep buffer. At each sampled sub-pixel location $(x,y,t)$, we find current spatial gradients by sampling four cotemporal image locations $(x_{\pm 1},y,t)$ and $(x,y_{\pm 1},t)$ each one pixel width from $x,y$, and then calculating the average horizontal and vertical absolute luminance differences $|L_{x,y,t} - L_{x \pm 1,y,t}|/2$ and $|L_{x,y,t} - L_{x,y \pm 1,t}|/2$. To find a current temporal gradient, we must compare two spatially collocated samples made at different times. We accomplish this by finding the absolute luminance difference between the center $(x,y,t)$ sample and a sample $(x,y,t_0)$ made previously at the same sub-pixel spatial location, and divide by the time elapsed since that previous sample was made:

```
fill deep buffers non-adaptively
loop
  update view and animation state to current time t
  choose a tile to render
  find last sample made in tile at sub-pixel location xyt0
  complete crosshair with 5 new samples centered at (x,y,t)
  update deep buffers and tile statistics
  repeat 5 times
    choose a tile crosshair and reproject it
    reevaluate gradients in crosshair
    check visibility of crosshair center sample
    if occluded then create new crosshair at same location
    update deep buffers and tile statistics
  end repeat
  choose a new sub-pixel location x,'y' in tile to sample
  initiate crosshair with sample at (x,'y,'t), set xyt0 to (x',y',t)
  update deep buffers and tile statistics
  if one refresh time elapsed
    then send reconstructor view and tile information
  if another chunk of crosshairs has been completed
    then adjust tiling
end loop
```

**Figure 5:** *Pseudocode for main loop in the sampler.*

$|L_{(x,y,t)} - L_{x,y,t0)}|/(t-t_0)$. Each crosshair is pushed into the *xy* queue of the deep buffer. To determine average gradients, we reduce the weight of each sample gradient as a function of age using the same exponential scheme used to track color variation.

### 4.2. Algorithm details and output to reconstructor

The complete sampling algorithm is described in Figure 5, including an inner reprojection loop which we will discuss later in Section 6.1. The deep buffer is initialized using a spatially gridded, cotemporal sampling pattern. Sampling iteration then begins and continues until the user halts rendering. Note that sampling of temporal gradients requires distributing crosshair sampling across two sampling iterations. In each iteration, one crosshair consisting of old samples at $(x,y,t_0)$ and new samples at $(x,y,t)$, $(x_{\pm1},y,t)$ and $(x,y_{\pm1},t)$ is completed, and another crosshair is initiated with a new sample at location $(x,'y,'t)$ (which will become old sample $(x,y,t_0)$ in the next iteration that visits the same tile). These initial crosshair samples are not stored in the tiling, but in a separate image-sized array, avoiding the complications caused by retiling before the crosshair is completed.

All samples in each crosshair are also immediately streamed to the reconstructor's deep buffer, without any of the crosshair structure. At each display refresh, the sampler also sends the current view and tiling to the reconstructor, including each tile's image coordinates as well as the average temporal and spatial gradients in the tile's block. The sampler's tiling is updated after a *chunk* (currently 25) of new crosshairs have been generated.

### 4.3. Comparison to prior work: sampling

Rather than a deep buffer, Tapestry [SS00] stores samples as vertices in a 2.5D Delaunay triangulated mesh spanning a sphere around the eyepoint. This mesh is used both in reconstruction (see Section 5.3 below) and as feedback for adaptive sampling. At the beginning of each frame, a priority image is formed by rendering the error assigned to each face in the mesh as color. Total face error increases with face area, the age of the samples forming its vertices, and the differences in sample color and depth among its vertices. The priority image's pixels are then traversed quasi-randomly in a space-filling fashion. Whenever error at a pixel rises above a certain threshold, that pixel is sampled and placed into the mesh, and mesh correctness checked.

Sampling in the Shading Cache [TPWG02] has several similarities to Tapestry. Rather than a 2.5 mesh, Shading Cache places samples into a subdividable 3D mesh that in fact represents the entire scene being rendered. Once more the mesh is rendered at each frame to form a priority image, which guides sampling. An additional flood fill stage hones in on high priority image discontinuities. Total error of each mesh face once more increases with face size, color differences, and age (on specular or moving objects).

The Render Cache [WDP99; WDG02] places samples into a fixed-size list. Priority images are generated with each frame, in which a pixel's sampling priority depends on the local sample density and age. Large color differences between consecutive samples in the same pixel age samples in nearby pixels more quickly. View prediction compensates for the undersampling caused by frame and network delays, e.g. at image edges during rotation.

Our renderer differs in its use of a frameless sampling pattern, which allows lower-latency response to scene change. To exploit this possibility, our sampler guides adaptive sampling with an incrementally and frequently updated tiling rather than a priority image formed once per frame. Error does not increase with sample age, but directly with temporal color variation, allowing temporal sampling density to drop drastically in static image regions (increasing average sample age), and increase steeply in regions containing motion (decreasing age). In previous methods the scene cannot change during a single frame, so these systems can form priority images with confidence. Since in our adaptive frameless renderer the scene can change after every crosshair is generated, we vary adaptive response (gain) using methods inspired by control engineering.

## 5. Interactive space-time reconstruction

Frameless sampling strategies demand a rethinking of the traditional computer graphics concept of an "image", since at any given moment the samples in an image plane represent many different moments in time. The original frameless work [BFMS94] simply displayed the most recent sample at every pixel. This *traditional reconstruction* results in a noisy image that appears to sparkle when the scene is dynamic (Figure 1). In contrast, we convolve the frameless samples in the reconstructor's deep buffer with space-time filters to continuously reconstruct images for display. This is similar to the classic computer graphics problem of reconstruction of an image from non-uniform samples [M87], but with a temporal element: since older samples may represent "stale" data, they are treated with less confidence and contribute less to nearby pixels than more recent samples (Figure 1).

### 5.1. Choosing a filter

The key question is what shape and size filter to use. A temporally narrow, spatially broad filter (i.e. a filter which falls off rapidly in time but gradually in space) will give very little weight to relatively old samples, emphasizing the newest samples and leading to a blurry but very current image. Such a filter provides low-latency response to changes and should be used when the underlying image is changing rapidly. A temporally broad, spatially narrow filter will give nearly as much weight to relatively old samples as to recent samples; such a filter accumulates the results of many samples and leads to a finely detailed, antialiased image when the underlying scene is changing slowly. However, often different regions of an image change at different rates, as for example in a stationary view in which an object is moving across a static background. A scene such as this demands adaptive reconstruction, with filter space-time extent varying across the image.

We use local sampling density (Figure 7) and space-time gradient information (Figure 4) to guide filter size. The reconstructor maintains an estimate of local sampling density across the image, based on the tiling used to guide sampling. We size our filter support—which can be interpreted as a space-time volume—as if we were reconstructing a regular sampling with this local sampling density, and while preserving the total volume of the filter, perturb the spatial and temporal filter extents according to local gradient information. A large spatial gradient implies an edge, which should be resolved with a narrow filter to avoid blurring across that edge. Similarly, a large temporal gradient implies a "temporal edge" such as an occlusion event, which should be resolved with a narrow filter to avoid including stale samples from before the event. This is equivalent to an "implicit" robust estimator; rather than searching for edges explicitly, we rely on the gradient to allow us to size the filter such that the expected contribution of samples past those edges is small.

Thus, given a local sampling rate $R_l$, expressed in samples per pixel per second, we define $V_S$ as the expected space-time volume occupied by a single sample:

$$V_S = \frac{1}{R_l}.$$

The units of $V_S$ are pixel-seconds per sample (note that the product of pixel areas and seconds is a volume). We then construct a filter at this location with space-time support proportional to this volume. For simplicity we restrict the filter shape to be axis-aligned to the spatial $x$ and $y$ and the temporal $t$ dimensions. The filter extents $e_x$, $e_y$, and $e_t$ are chosen to span equal expected color change in each dimension, determined by our estimates of the gradients $G_x$, $G_y$, and $G_t$ and the total volume constraint $V_s$:

$$e_x G_x = e_y G_y = e_t G_t$$

$$V_S = e_x e_y e_z$$

Thus the filter extents are given by

$$e_x = \sqrt[3]{\frac{V_S G_y G_t}{G_x^2}}, e_y = \sqrt[3]{\frac{V_S G_x G_t}{G_y^2}}, e_t = \sqrt[3]{\frac{V_S G_x G_y}{G_t^2}}.$$

What function to use for the filter kernel remains an open question. We have experimented with a range of filters. The Mitchell-Netravali filter [M87] is considered among the best filters for nonuniform sampling, but is costly and requires more precision than the 16-bit floating point buffers of our GPU implementation provide. We have also experimented with a simple inverse exponential filter, which has the nice temporal property that the relative contribution of two samples does not change as both grow older; however, the bandpass properties of this filter are less than ideal. We currently use a Gaussian filter.

### 5.2. Scatter versus gather

We can consider reconstruction a *gather* process which loops over the pixels, looks for samples in the neighborhood of each pixel, and evaluates the contribution of those samples to that pixel. Alternatively, we can cast reconstruction as a *scatter* process which loops over the samples, projects each onto the image plane, and evaluates its contribution to all pixels within some footprint. We have experimented with both approaches.

We implemented the reconstructor initially as a gather process directly on the sampler's deep buffer. At display time the reconstructor looped over the pixels, adjusting the filter size and extents at each pixel using gradient and local sample density as described above. The reconstructor gathered samples outwards from each pixel in space and time until the maximum possible incremental contribution of additional samples would be less than some threshold $\varepsilon$. The final color at that pixel was computed as the normalized weighted average of sample colors. This process proved expensive in practice—our unoptimized simulator required reconstruction times of several hundred ms for small (256 × 256) image sizes. It was also unclear how to efficiently implement hardware sample reprojection.

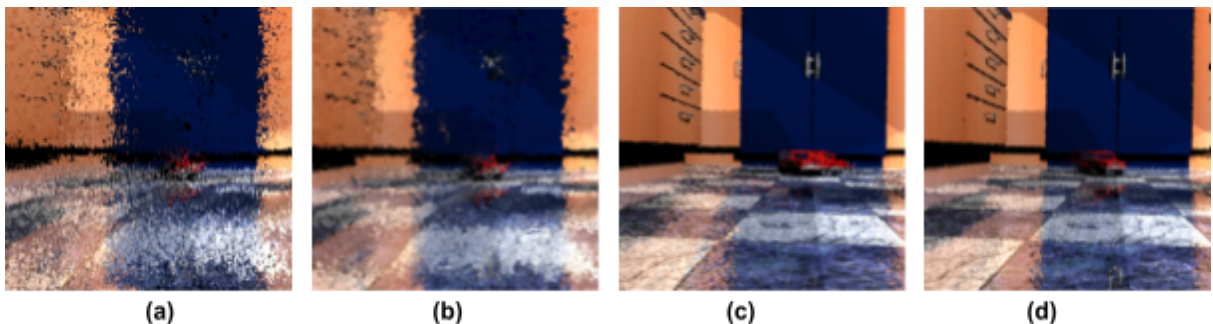

**Figure 6:** *Adaptive reconstruction illustrated in one moment of a scene with a moving view and car, sampled using our adaptive frameless techniques. In (a), traditional frameless reconstruction leaves many artifacts of the view motion in the image. In (b), adaptive reconstruction rejects many of the outdated samples, eliminating artifacts and clarifying edges. (c) shows the improvements possible by reprojecting samples as in [WDP99], even without adaptive reconstruction. When reprojection is combined with adaptive reconstruction as in (d), the car's motion and view-dependent reflections in the floor are clarified.*

We have therefore moved to a scatter-based implementation that stores the $N$ most recent samples produced by the sampler across the entire image; the value of $N$ is typically at least 4× the desired image resolution. This store is a distinct deep buffer for the reconstructor that organizes the samples as a single temporally ordered queue rather than a spatial array of crosshairs. At reconstruction time, the system splats each of these samples onto the image plane and evaluates the sample's affect on every pixel within the splat extent by computing the distance from the sample to the pixel center and weighting the sample's color contribution according to the local filter function. These accumulated contributions are then divided by the accumulated weight at each pixel to produce the final image (Figure 6).

We implement this scatter approach on the GPU, improving on the speed of our CPU-based gather implementation by almost two orders of magnitude. The GPU treats the samples in the deep buffer as vertices in a vertex array, and uses an OpenGL vertex program to project them onto the screen as splats (i.e., large GL_POINTS primitives). A fragment program runs at each pixel covered by a sample splat, finding the distance to the sample and computing the local filter shape by accessing tile information (local filter extent and $G_x, G_y, G_t$ gradients) stored in a texture. This texture is periodically updated by rasterizing the latest tiling (provided by the sampler) as a set of rectangles into an offscreen buffer. To reduce overdraw while still providing broad filter support in sparsely sampled regions, the vertex program rendering the samples adaptively adjusts point size (see Section 6.2).

The reconstructor uses several features of recent graphics hardware, including floating-point textures with blend support, multiple render targets, vertex texture fetch, dynamic branching in vertex programs, and separate blend functions for color and alpha. The results presented in this paper were obtained on a NVIDIA GeForce 6800 Ultra, which can reconstruct a visually sufficient number of samples for 256×256 pixel imagery ($N$=400K) at about 20 Hz.

### 5.3. Comparison to prior work: reconstruction

Both Tapestry [SS00] and the Shading Cache [TPWG02] use graphics hardware to perform a piecewise linear reconstruction of sparse samples by incorporating the samples as vertices into a Gouraud-shaded mesh, which is then rendered normally. In the case of Tapestry, this mesh is a Delaunay triangulation of a height field surrounding the viewpoint; in the case of the Shading Cache, it is a hierarchical subdivision meshing of the actual 3D scene similar to those used in radiosity methods. In both cases maintenance of the mesh requires a relatively expensive remeshing step that by its nature must happen on the CPU, though the Shading Cache amortizes this cost across several frames by double-buffering the display mesh and the mesh being updated.

Both techniques target extremely sparse sampling rates, citing examples in the range of 50-400 samples per frame.

By design, a mesh-based reconstruction gives equal weight to all samples; the color of each pixel is fully determined by the three samples that form the vertices of the triangle visible at that pixel. The use of Z-buffered rasterization precludes incorporating multiple samples with different weights, for example to incorporate our temporal filters or to perform antialiasing. While Tapestry does not require the original 3D model or place any restrictions on its geometry, Shading Cache utilizes the full geometry of the original scene, which it requires in a locally parameterizable form suitable for hierarchical subdivision. As a result it provides very crisp imagery with no reprojection artifacts, but may not scale well to very complex scenes.

The Render Cache [WDP99; WDG02] uses an image-based reconstruction closer in spirit to our work. Two fixed-size image-space kernels are used: a 3×3 Gaussian filter reconstructs most pixels, and a 7×7 box "prefilter" provides coarse hole-filling where the 3×3 filter cannot reach any samples, i.e. for heavily disoccluded regions. The principal difference in our work is the use of filters with a temporal aspect, whose size and shape are fully adaptive and vary across the image as described. The Render Cache also uses a depth-sensitive filter that helps prevent occluded samples from "leaking through" the occluding surfaces, while we rely on the sampler to detect such occlusion artifacts (and the high temporal gradients they create) and to respond by sending more samples and emphasizing recent samples representing the correct occlusion.

## 6. Reprojection

Our adaptive frameless sampling and reconstruction techniques operate entirely in 2D image space and do not rely on information about sample depth or the 3D structure of the scene. However, because camera location and sample depth are easily available from our ray-tracing renderer, we also incorporate sample reprojection [WDP99; WDG02; BDT99; WS99] into our algorithms. During sampling, reprojection can help the sampler find and focus on image regions undergoing disocclusion, occlusion, view-dependent lighting changes, or view-independent motion. During reconstruction, sample reprojection extends the effective "lifetime" of a sample by allowing older samples to contribute usefully to imagery even after significant camera or object motion. We use different strategies for reprojection within the sampler and reconstructor.

### 6.1. Reprojection in the sampler

It is not necessary to reproject every sample at fixed intervals, and indeed this would not be desirable since it would introduce periodicity (i.e., frames) into our frameless sampler. Instead, we reproject a small number of recent samples as we generate each new sample. Reprojection is quite fast; when updates of tiling statistics (e.g. variation, gradients) are included, reprojecting a sample takes roughly $1/35^{th}$ the mean time required to generate a new sample in our test datasets. We therefore reproject a small number $r$ (currently $r=5$) of crosshairs from a tile each time the sampler visits that tile to generate a new sample. In this way the same rendering bias that guides generation of new samples also guides reprojection of existing samples, focusing reprojections on important image areas.

When the sampler visits a tile, it chooses $r$ pixels to reproject randomly and relocates the crosshairs from the front of each pixel's queue in the deep buffer (Figure 5). To relocate a crosshair we apply both the current viewing transformation and stored sample velocity (if any) to the samples in the crosshair. We determine a crosshair's new location in the buffer by its relocated center sample, insert the crosshair at the back of its new queue, and update source and destination tile statistics if necessary. When updating tile gradients, spatial gradients for the crosshair are recalculated using the new spatial locations of the crosshair samples; we recalculate the crosshair temporal gradients by finding the absolute difference between the reprojected center sample and the newest sample in that pixel region, and dividing this difference by the age of this newest sample. Note that a crosshair may potentially be reprojected more than once.

Regions containing disocclusions will be undersampled as samples reproject to other image locations. We bias sampling toward these disocclusions with a new undersampling measure $u_{tile}$:

$$u_{tile} = 1 - \min\left(1, \frac{m \sum_{j=1}^{|tiles|}(whb - |buffer|)/|tiles|}{sb - |tile|}\right).$$

Here the number of empty samples in a tile must be $m$ times greater than the mean number of empty samples in all tiles to affect sampling. $|buffer|$ and $|tile|$ are the number of samples in the deep buffer and the current tile's block, while $whb$ is the number of samples the deep buffer can hold (with image size $w \times h$).

Regions undergoing occlusion will contain samples from multiple surfaces at differing view depths, leading to uncertainty about image content. To resolve this uncertainty, we increase sampling in occluded regions. We detect occlusions by casting rays from the center sample of each reprojected crosshair to the eye. Like shadow rays, these rays need no shading and can terminate if they hit any geometry at all. If this sample is no longer visible from the eye, we replace the reprojected crosshair with a new one centered at the same image location. We also increase sampling density in the occluded region by increasing error in tiles experiencing occlusion with an occlusion term $o_{tile} = |O|/sb$, where $|O|$ is the number of occluded samples in a tile's block, tracked by our occlusion test. Figure 7 shows the occlusions that affect sampling. Tile error $E_{tile}$ then becomes:

$$E_{tile} = s\left(\kappa \frac{v_{tile}}{\sum_j^{|tiles|} v_j} + \lambda \frac{u_{tile}}{\sum_j^{|tiles|} u_j} + (1-\kappa-\lambda)\frac{o_{tile}}{\sum_j^{|tiles|} o_j}\right).$$

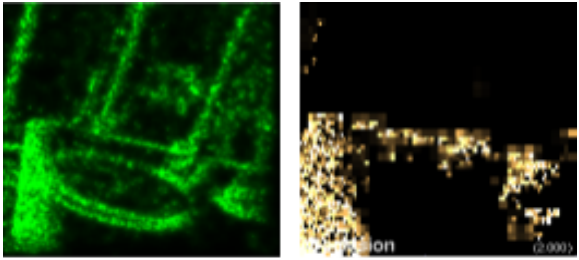

**Figure 7:** *A sample density map (left) used by the reconstructor to determine the expected local sample volume $V_s$, and occlusion detection (right) used direct sampling.*

Here $\kappa$, $\lambda$, and $(\kappa + \lambda)$ are all in [0,1]; $v_{tile}$, $u_{tile}$ and $o_{tile}$ are normalized by their sum over all tiles; and $s$ is tile size.

### 6.2. Reprojection in the reconstructor

Unlike the sampler, the reconstructor operates in a framed context: to display an image on existing hardware, it scans out a traditional image (i.e., a uniform grid of pixels) at the regular intervals of the display refresh. Since each sample in the reconstructor's deep buffer stores the 3D hit point of the primary ray that generated that sample, reprojecting and reconstructing each of our renderer's images reduces to rendering the vertex array in the deep buffer with the current camera and projection matrices bound. Figure 6 shows the results of using reprojection in reconstruction.

Reprojection can generate regions of low sample density, for example at disocclusions and near leading screen edges during view rotation. In such regions, filter support for the few samples present must be quite large, requiring rasterization of samples with large splats. Rather than rasterizing all samples with large splats, we adjust splat size adaptively. Samples are accumulated into a *coverage map* during rendering that tracks the number and average splat size of all samples rendered to each pixel. To size splats, the sample vertex program binds the previous image's coverage map as a texture, computes the projected coordinates of the sample, and uses the coverage information at those coordinates to calculate the splat size at which the sample will be rasterized. Sample splats in a region have the average size of splats used in that region during reconstruction of the previous image, but splat sizes in undersampled regions (defined currently as fewer than 4 samples affecting a pixel) are multiplied by 4 to grow rapidly, while splat sizes in oversampled regions (more than 32 samples reaching a pixel) are multiplied by 0.7 to shrink gradually.

### 7. Evaluation

Using the *gold standard* validation described in [WLWD03], we find that our adaptive frameless renderer outperforms traditional framed and frameless renderers using the same sampling rates. The control machinery that achieves this improved performance requires less than 15% of total computation time.

Gold standard validation uses as its standard a simulated ideal renderer capable of rendering antialiased imagery in zero time. To perform comparisons to this standard, we create $n$ ideal images $I_j$ at 60 Hz for an animation using a simulated ideal renderer and $n$ images $R_j$ for the same animation using an actual interactive renderer $R$. We next compare each image pair $(I_j, R_j)$ using an image comparison metric. Here we use root-mean-squared error (RMS).

| Render Method | Animation/Sampling rate | | | | | | |
|---|---|---|---|---|---|---|---|
| | Interactive | | | Bart | | Toycar | |
| | 100k | 400k | 800k | 400k | 800k | 400k | 800k |
| 400K 60Hz | 92.7 | 71.8 | 60.9 | 112 | 100 | 47 | 42.8 |
| 400K Full Res | 110 | 72.6 | 60.4 | 127 | 112 | 43.8 | 38.9 |
| Traditional Frameless | 80.8 | 48.8 | 39.3 | 92.3 | 74.8 | 35.3 | 32.5 |
| Adaptive | 34.4 | 24.1 | 23.6 | 50.1 | 51.9 | 20.4 | 18.5 |
| Full Res 60Hz | 28 | 28 | 28 | 30.7 | 30.7 | 29.4 | 29.4 |

**Table 1:** *Summary error analysis using the techniques of Figure 8, with some additional sampling rates.*

We report the results of our evaluation in Table 1, which compares several rendering methods producing 256x256 images using various sampling rates. Two framed renderings either maximize temporal resolution (i.e., frame rate) at the cost of spatial resolution (*400K 60Hz*), or maximize spatial resolution at the cost of temporal resolution (*400K Full Res*). The *traditional frameless* rendering simply displays the newest sample at a given pixel. The *adaptive* rendering uses our system to produce the imagery. *Full Res 60Hz* is a framed renderer that uses a sampling rate 10 times higher than all of the other renderers to produce full resolution imagery at 60Hz. (The difference between the ideal renderer and the *Full Res 60Hz* renderer is that the latter suffers from double-buffering delay and does not use anti-aliasing). Rendering methods were tested in 3 different animations, all using the publicly available BART testbed [LAM00]: the testbed viewpoint animation (*Bart*); a fixed viewpoint close-up of a moving car (*Toycar*), and a recording of user viewpoint interaction (*Interactive*).

Adaptive frameless rendering is the clear winner, with lower RMS error than all techniques using the same sampling rate and comparable error to the Full Res 60Hz rendering, with sampling rate 40, 10 and 5 times higher than the 100K, 400K and 800K adaptive frameless renderings.

Figure 8 offers a more detailed view. The graphs here show frame-by-frame RMS error comparisons between several of these rendering techniques and the ideal rendering. Note the sawtooth pattern produced by the *400K Full Res* renderer, due to double buffering error. In the interactive animation, the periodic increases in error correspond to periods of rapid viewpoint change. Again, adaptive frameless rendering has lower RMS error than all rendering techniques using equivalent sampling rates, and comparable error to the much more densely sampled *Full Res 60Hz* renderer. The top right graph also depicts the advantage of using reprojection in the sampler (*Adaptive no reprojections*). Error is considerably higher without reprojection.

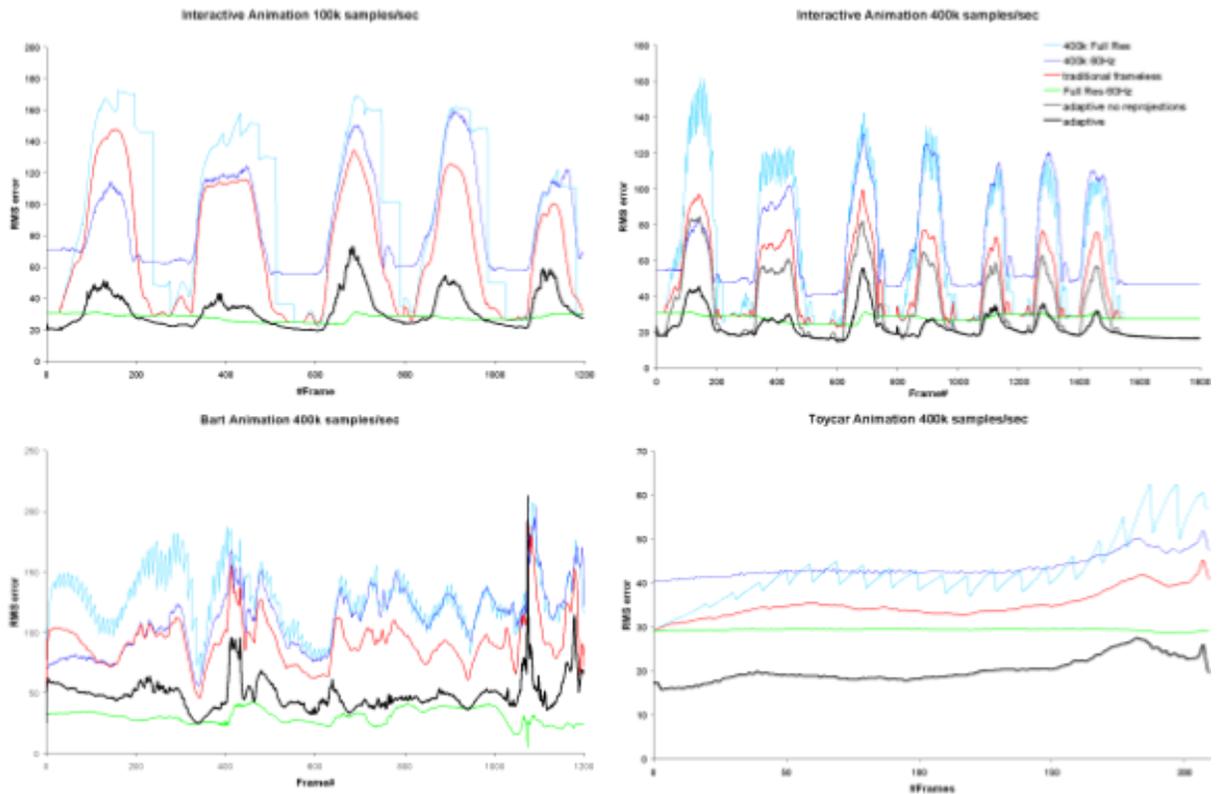

**Figure 8:** *Error analysis of rendering techniques for several animation sequences created using 100K or 400K samples/sec. Graphs show frame-by-frame RMS error between each technique's images and the ideal image that would be displayed by a hypothetical zero-delay, antialiased renderer at the same moment in time. Resolution is 256x256 pixels at 60 Hz.*

### 8. Discussion and future work

Frameless rendering and selective sampling have been criticized for sacrificing spatial coherence and thus memory locality, which can reduce sampling speed. We will experiment with increases in the number of samples we generate each time we visit a tile, increasing spatial coherence at the cost of slightly less adaptive sampling overall. However, exploiting spatial coherence has its limits: ultimately, it will limit our ability to take advantage of temporal coherence and force us to sample more often. Traditional renderers must sample every single pixel dozens of times each second; as displays grow in size and resolution, this ceaseless sampling becomes wasteful of computation, power, and heat. With this work, we hope to shift the emphasis of interactive ray tracing research from spatial to temporal coherence, and from brute-force to selective sampling.

Good filter design for adaptive space-time reconstruction of frameless sample streams remains an open problem. We have begun investigating the edge-preserving concepts of bilateral and trilateral filtering [DD02; CT03], which perform nonlinear filtering by weighting samples according to their difference in luminance as well as their distance in space. However, extending these approaches to include a third temporal dimension and to operate on non-uniformly distributed samples presents a significant challenge. Another possibility is to exploit a priori information about the underlying model or animation, as do Bala et al. [BWG03].

We will continue this research in several longer-term directions. Extending our temporally adaptive methods to more sophisticated global illumination algorithms is one obvious avenue. With its ability to selectively alter sampling and reconstruction across both space and time, our adaptive frameless renderer is an ideal platform for experimenting with perceptually driven rendering in interactive settings [LRC02]. We are studying the possibility of extremely high resolution ("gigapixel") display hardware fed streams of frameless samples, with adaptive reconstruction performed in the display itself. This might be one solution to the immense bandwidth challenge posed by such displays. Such a rendering configuration would also enable a truly asynchronous parallelism in graphics, since renderers would no longer have to combine their samples into a single frame. For this reason we are particularly interested in implementing these algorithms in graphics hardware.

## 9. Conclusion

In conclusion, we advocate a revival of frameless rendering, enabled by recent advances in interactive ray tracing and based on spatio-temporally adaptive sampling and reconstruction. Adaptive frameless rendering incorporates techniques from adaptive renderers, reprojecting renderers, non-uniform reconstruction, and GPU programming. The resulting system outperforms traditional framed and traditional frameless renderers and offers the following extensions of previous reprojecting renderers:

*Improved sampling response.* Rather being clustered at each frame time, samples reflect the most up-to-date input available at the moment they are created. Closed-loop control guides samples toward not only spatial but temporal color discontinuities at various scales. These elements combine to reduce rendering latency.

*Improved reconstruction.* Rather than being non-adaptive or hardware-interpolated, reconstruction is adaptive over both space and time, responding to local space-time color gradients. This significantly improves image quality, eliminating the temporal incoherence in traditional frameless imagery without requiring framed sampling and its increased latency, and permitting antialiasing in static image regions. Reconstruction is implemented on existing GPU hardware.

Our prototype system displays greater accuracy than framed and frameless rendering schemes at comparable sampling rates, and similar accuracy to a framed renderer sampling 10 times more quickly. Based on these results, we believe that adaptive frameless approach shows great promise for future rendering algorithms and hardware.

## 10. Acknowledgements

Thanks to Greg Humphreys, who created the video, and to Bruce Walter for discussions of the Render Cache. Our gratitude to Ed Colgate and Kevin Lynch for their discussions of control engineering. This research was supported by NSF grants 0092973, 0093172, 0112937, and 0130869.

## 11. References

[BDT99] BALA, K., DORSEY, J., TELLER, S. 1999. Radiance interpolants for accelerated bounded-error ray tracing. *ACM Trans. Graph*, 18, 3, 213-256.

[BWG03] BALA, K., WALTER, B., GREENBERG, D.P. 2003. Combining edges and points for interactive high-quality rendering. *ACM Trans. Graph., 22*, 3, 631–640 (*Proc. ACM SIGGRAPH*).

[BFGS86] BERGMAN, L., FUCHS, H., GRANT, E., SPACH, E. 1986. Image rendering by adaptive refinement. *Proc. ACM SIGGRAPH*, 29–37.

[BFMS94] BISHOP, G., FUCHS, H., MCMILLAN, H., SCHER ZAGIER, E.J. 1994. Frameless rendering: double buffering considered harmful. *Proc. ACM SIGGRAPH*, 175–176.

[CHH02] CARR, N.A., HALL, J.D., HART, J.C. 2002. The ray engine. *Proc. ACM SIGGRAPH/Eurographics Graphics Hardware*, 37–46.

[CT03] CHOUDHURY, P., TUMBLIN, J. 2003. The trilateral filter for high contrast images and meshes. *Proc. Eurographics Workshop on Rendering*, 186–196.

[DD02] DURAND, F., DORSEY, J. 2002. Fast bilateral filtering for the display of high-dynamic-range images. *ACM Trans. Graphics, 21*, 3, 257–266 (*Proc. ACM SIGGRAPH*).

[G95] GLASSNER, A. 1995. *Principles of Digital Image Synthesis, 1st ed*. Morgan Kaufmann.

[HDM03] HAVRAN, V., DAMEZ, C., MYSZKOWSKI, K. 2003. An efficient spatio-temporal architecture for animation rendering. *Proc. Eurographics Symposium on Rendering*, 106-117.

[J01] JENSEN, H.W. 2001. *Realistic Image Synthesis Using Photon Mapping*. AK Peters.

[LAM00] LEXT, J., ASSARSSON, U., MOELLER, T. 2000. Bart: A benchmark for animated ray tracing. Tech. Rpt. 00-14, Dept. Computer Engineering, Chalmers Univ. Tech. http://www.ce.chalmers.se/BART.

[LRC*02] LUEBKE, D., REDDY, M., COHEN, J.D., VARSHNEY, A., WATSON, B., HUEBNER, R. 2002. *Level of Detail for 3D Graphics, 1st ed*. Morgan Kaufmann.

[M87] MITCHELL, D.P. 1987. Generating antialiased images at low sampling densities. *Proc. ACM SIGGRAPH*, 65–72.

[OCMB95] OLANO, M., COHEN, J., MINE, M., BISHOP, G. 1995. Combatting rendering latency. *Proc. ACM Interactive 3D Graphics*, 19–24.

[PS89] PAINTER, J., SLOAN, K. 1989. Antialiased ray tracing by adaptive progressive refinement. *Proc. ACM SIGGRAPH*, 281–288.

[PMS*99] PARKER, S., MARTIN, W., SLOAN, P.-P.J., SHIRLEY, P., SMITS, B., HANSEN, C. 1999. Interactive ray tracing. *Proc. ACM Interactive 3D Graphics*, 119–126.

[PKGH97] PHARR, M., KOLB, C., GERSHBEIN, R., HANRAHAN, P. 1997. Rendering Complex Scenes with memory-coherent ray tracing. *Proc. ACM SIGGRAPH*, 101–108.

[PBMH02] PURCELL, T.J., BUCK, I., MARK, W.R., HANRAHAN, P. 2002. Ray tracing on programmable graphics hardware. *ACM Trans. Graphics, 21*, 3, 703–712 (*Proc. ACM SIGGRAPH*).

[RP94] REGAN, M.J.P., POSE, R. 1994. Priority rendering with a virtual reality address recalculation pipeline. *Proc. ACM SIGGRAPH*, 155–162.

[RSH05] RESHETOV, A., Soupikov, A., Hurley, J. 2005. Multi-Level Ray Tracing Algorithm. *ACM Trans. Graph., 24*, 3, (*Proc. ACM SIGGRAPH, to appear Aug 2005*).

[S97] SCHER-ZAGIER, E. 1997. *Defining and Refining Frameless Rendering*. University of North Carolina Technical Report #TR97-008.

[SWWPS04] SCHMITTLER, J., WOOP, S., WAGNER, D., PAUL, W., and SLUSALLEK, P. 2004. Realtime Ray Tracing of Dynamic Scenes on an FPGA Chip. *Proc. Graphics Hardware 2004*.

[SS00] SIMMONS, M., SÉQUIN, C. 2000. Tapestry: A dynamic mesh-based display representation for interactive rendering. *Proc. Eurographics Workshop on Rendering*, 329–340.

[SHSS00] STAMMINGER, M., HABER, J., SCHIRMACHER, H., and SEIDEL, H. 2000. Walkthroughs with Corrective